\documentclass[aps,pra,notitlepage,reprint,superscriptaddress,
nofootinbib,longbibliography]{revtex4-1}

\usepackage{graphicx}
\usepackage{dcolumn}
\usepackage{bm}
\usepackage{amsmath,amssymb}
\usepackage{upgreek}
\usepackage{mathrsfs}

\usepackage{hyperref}
\hypersetup{colorlinks,linkcolor=red,citecolor=blue,urlcolor=red}
\usepackage{pdfpages}
\usepackage{tikz}
\usepackage[utf8]{inputenc}
\usepackage[T1]{fontenc}
\usepackage{mathptmx}
\usepackage{eucal}
\usepackage[normalem]{ulem}
\usepackage[mathlines]{lineno}
\usepackage{comment}
\renewcommand{\t}[1]{\mathrm{#1}}
\DeclareMathAlphabet{\mathcal}{OMS}{cmsy}{m}{n}

\makeatletter
\AtBeginDocument{\let\LS@rot\@undefined}
\makeatother

\AtBeginDocument{%
	\newwrite\bibnotes
	\def\bibnotesext{Notes.bib}
	\immediate\openout\bibnotes=\jobname\bibnotesext
	\immediate\write\bibnotes{@CONTROL{REVTEX41Control}}
	\immediate\write\bibnotes{@CONTROL{%
			apsrev41Control,author="08",editor="1",pages="1",title="0",year="1"}}
	\if@filesw
	\immediate\write\@auxout{\string\citation{apsrev41Control}}%
	\fi
}%
\begin{document}
    
\title{Optimum classical beam position sensing}


\author{Wenhua He}
\affiliation{Wyant College of Optical Sciences, University of Arizona, Tucson, AZ 85721, USA}

\author{Christos N. Gagatsos}
\affiliation{Department of Electrical and Computer Engineering, University of Arizona, Tucson AZ 85721}

\author{Dalziel J. Wilson}%
\affiliation{Wyant College of Optical Sciences, University of Arizona, Tucson, AZ 85721, USA}

\author{Saikat Guha}%
\affiliation{Wyant College of Optical Sciences, University of Arizona, Tucson, AZ 85721, USA}

\date{\today}
\begin{abstract}
Beam displacement measurements are widely used in optical sensing and communications; however, their performance is affected by numerous intrinsic and extrinsic factors including beam profile, propagation loss, and receiver architecture.  Here we present a framework for designing a classically optimal beam displacement transceiver, using quantum estimation theory.   We consider the canonical task of estimating the position of a diffraction-limited laser beam after passing through an apertured volume characterized by Fresnel-number product $D_\text{F}$.  As a rule of thumb, higher-order Gaussian modes provide more information about beam displacement, but are more sensitive to loss.  Applying quantum Fisher information, we design mode combinations that optimally leverage this trade-off, and show that a greater than 10-fold improvement in precision is possible, relative to the fundamental mode, for a practically relevant $D_\text{F}=100$.  We also show that this improvement is realizable with a variety of practical receiver architectures.  Our findings extend previous works on lossless transceivers, may have immediate impact on applications such as atomic force microscopy and near-field optical communication, and pave the way towards globally optimal transceivers using non-classical laser fields.
\end{abstract}

\maketitle

Estimating the transverse displacement of a laser beam is a key task in a broad range of commercial and scientific applications, from atomic force microscopy~\cite{putman1992afm} and single-molecule tracking~\cite{taylor2013_tracking} to pointing-acquisition-tracking for free-space optical communications \cite{kaymak2018FSOsurvey} and telescope  stabilization~\cite{jwst_control_operation}. Extensive theoretical and experimental work has been dedicated to realizing improved transceiver designs. In most of these studies, the transmitter is the fundamental Hermite-Gaussian mode, $\t{HG}_{00}$---both in the classical regime, where the laser is in a coherent state, and in quantum-enhanced schemes (e.g., by mixing a $\t{HG}_{00}$ coherent state with phase-inverted $\t{HG}_{00}$ squeezed vacuum~\cite{2000-SD-theory, 2002-SD-EXP1D,2003-SD-EXP2D}).  Common receiver architectures include the split photodetector \cite{2000-SD-theory, 2002-SD-EXP1D,2003-SD-EXP2D}, lateral effect photodiode \cite{LEPD2022}, and homodyne interferometers employing phase-inversion \cite{larson2020SLIVER} or structured local oscillators \cite{2006-d-t-HG10-theory-exp,2006-d-t-HG10,2014-QFI-HGn0}.

In designing beam position transceivers that move beyond conventional $\t{HG}_{00}$ transmitters, a key insight is that higher order Gaussian modes provide more information about beam position (via their high spatial frequency content) \cite{2006-d-t-HG10,2014-QFI-HGn0,2006-d-t-HG10-theory-exp,qi2018ultimate}, albeit at the price of higher sensitivity to loss \cite{shapiro2005ultimate}.  Transceivers employing high order Hermite-Gaussian (HG) modes have been experimentally studied~\cite{2006-d-t-HG10,2014-QFI-HGn0,2006-d-t-HG10-theory-exp}, but only for a single higher order mode. Classically optimal transmitters employing single higher-order HG modes have also been theoretically considered~\cite{2014-QFI-HGn0,qi2018ultimate} using single parameter estimation theory~\cite{braunstein1994QFI}; however, as simplifying assumptions, both diffraction and loss in these studies were ignored.

 \begin{figure}[t!]
\centering  \includegraphics[width=0.7\columnwidth]{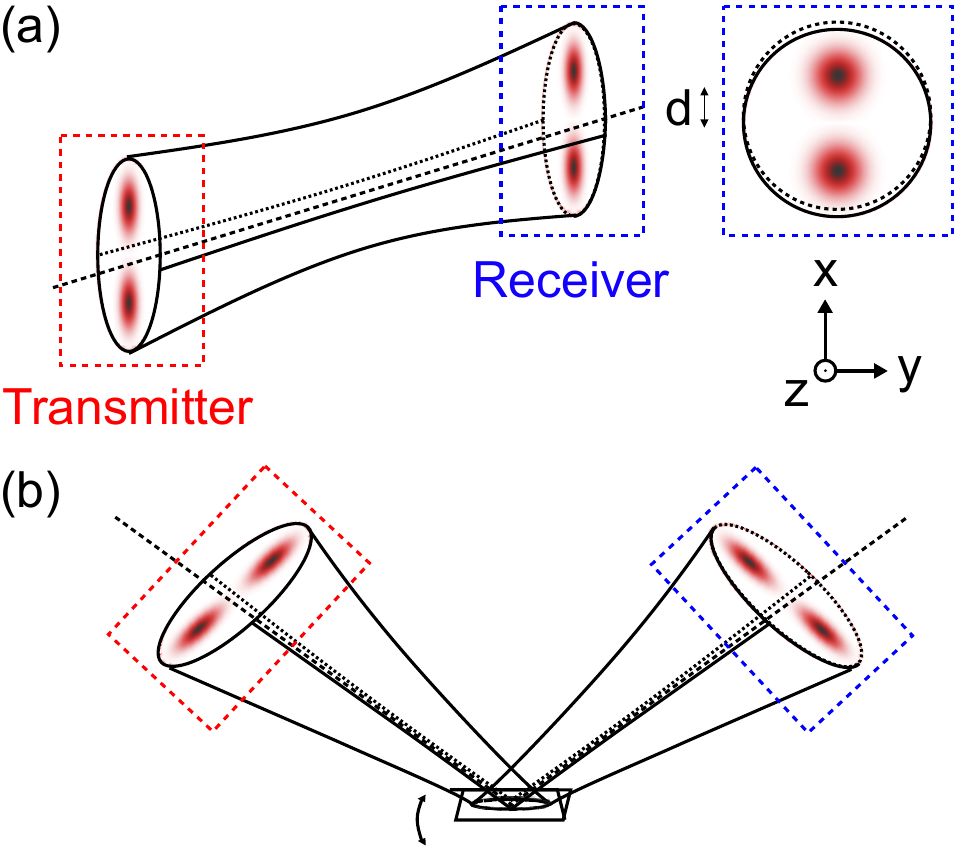}
\caption{(a) Transceiver model for optical beam displacement sensing.  Diffractive loss is introduced via the finite aperture of the transmitter and receiver planes, characterized by Fresnel-product number $D_\t{F}$.  A transverse intensity pattern that maximizes the Quantum Fisher Information for $D_\t{F} = 90$ is shown, assuming a coherent state with a specific energy. (b) Application to measuring the angular displacement of a reflective landscape (an optical lever measurement).}
\label{Fig_1}
\vspace{-2mm}
\end{figure}

Here, we present a framework for designing a classically optimal beam displacement transceiver that allows for diffraction, loss, and arbitrary spatial modeshape, based on quantum estimation theory.  The enabling tool for our study is quantum Fisher information (QFI) \cite{braunstein1994QFI, braunstein1996generalized}, which allows the spatial mode to be optimized, for a given laser (probe) state, over all possible receivers. Thus we are able to model the generic problem illustrated in Fig.\ref{Fig_1}a, in which a Gaussian laser beam passes through an apertured (at the transmitter and receiver plane) volume characterized by Fresnel number product $D_\t{F}$.  
After identifying an (in general non-unique) optimal spatial mode, the corresponding optimal receiver can then be found using classical Fisher information (CFI), 
a procedure which also yields a variety of options.

As a practical illustration \cite{SLM-structured-light,SLMbeyongdisplay,phc_SLM}, we confine our attention to coherent state transmitters and soft Gaussian apertures, relevant to common beam displacement transceivers such as the optical lever (Fig. \ref{Fig_1}b). For practically relevant Fresnel number product $D_\t{F} < 100$ (approximately the number of HG modes that will survive propagation through the system), we ``discover'' a class of bimodal transmitter modes as shown in Fig. \ref{Fig_1}a, which represent the transverse intensity (and phase) distribution that optimally weights the tradeoff between spatial derivative and diffraction loss.  For $D_\t{F} = 100$, we predict that such a mode can outperform the $\t{HG}_{00}$ mode by an order of magnitude, using a variety of standard receivers.

The starting point for our analysis is an abstract Fisherian description of beam displacement sensing.  For a given Fresnel number product $D_\t{F}$, transmitter mode, and receiver architecture, the CFI,  $J(d)$, upper bounds the inverse of the minimum uncertainty with which the laser beam displacement $d$ can be estimated 
~\cite{vantreebook,braunstein1994QFI}. 
The QFI, $K(d)$, is similarly related to $d$, but 
optimized over all possible receivers \cite{braunstein1994QFI, braunstein1996generalized}, and therefore a property of the transmitter state and $D_\text{F}$ alone, viz.
\begin{equation}\label{eq:1}
\frac{1}{\Delta d}\le {\sqrt{J(d)}} \le {\sqrt{K(d)}}.
\end{equation}
We thus adopt a mechanistic approach whereby $K(d)$ is first optimized, yielding a subspace of optimal transmitter modes.  The subset of optimal receiver architectures is then inferred by inspection, appealing to the upper bound set by Eq. \ref{eq:1}. 


To compute $K(d)$, we seek an expression for the multimode coherent state $|\vec{a}(d)\rangle$ at the receiver, and use the fact that 
\begin{equation}
K(d) = -2 \lim_{\epsilon\to 0}\frac{\partial^2\mathcal{F}}{\partial \epsilon^2},
\end{equation}
where 
\begin{equation}
\mathcal{F}(\epsilon) = |\langle \vec{a}(d)|\vec{a}(d+\epsilon)\rangle|^2,
\end{equation}
is the fidelity \cite{uhlmannfidelity} between states with different displacements.  Following a standard approach, $|\vec{a}(d)\rangle$ can be represented as a vector of mean values $\{\langle \hat{q}_0\rangle, \langle \hat{q}_1\rangle,... \langle \hat{p}_0\rangle, \langle \hat{p}_1\rangle,...\}$, where $\hat{a}_n$, $\hat{q}_n = \hat{a}_n+\hat{a}_n^\dagger$ and $\hat{p}_n = -i(\hat{a}_n-\hat{a}_n^\dagger)$ are the annihilation, position and momentum operators for field at the receiver, decomposed into an orthonormal basis of spatial modes 
\cite{weedbrook2012gaussian}.  It thus suffices to determine the modal decomposition of the laser field at the receiver and its functional dependence on $d$.

We consider a coherent state of mean photon number $N$, and choose as our orthonormal basis the HG modes---$\Phi_n(x)\Phi_m(y)$ at the transmitter and  $\phi_n(x)\phi_m(y)$ at the receiver (Fig. \ref{Fig_1})---as they form a singular value decomposition of the soft-aperture propagation kernel \cite{shapiro2005ultimate, SI}. 
Noting that only $\Phi_n(x)$ is sensitive to displacement and $\Phi_0(y)$ is least sensitive to diffraction loss, an optimal transmitted field can be expressed as 
\begin{equation}
   \Psi(x,y)= \sqrt{N}\sum\displaylimits^{M_\t{s}-1}_{n=0} c_n e^{i \theta_n} \Phi_n(x)\Phi_0(y),
\end{equation}
with expansion coefficients $c_n,\theta_n\in \Re$ satisfying $\sum_n c_n^2=1$.

 \begin{figure}[t!]
\centering  \includegraphics[width=1\columnwidth]{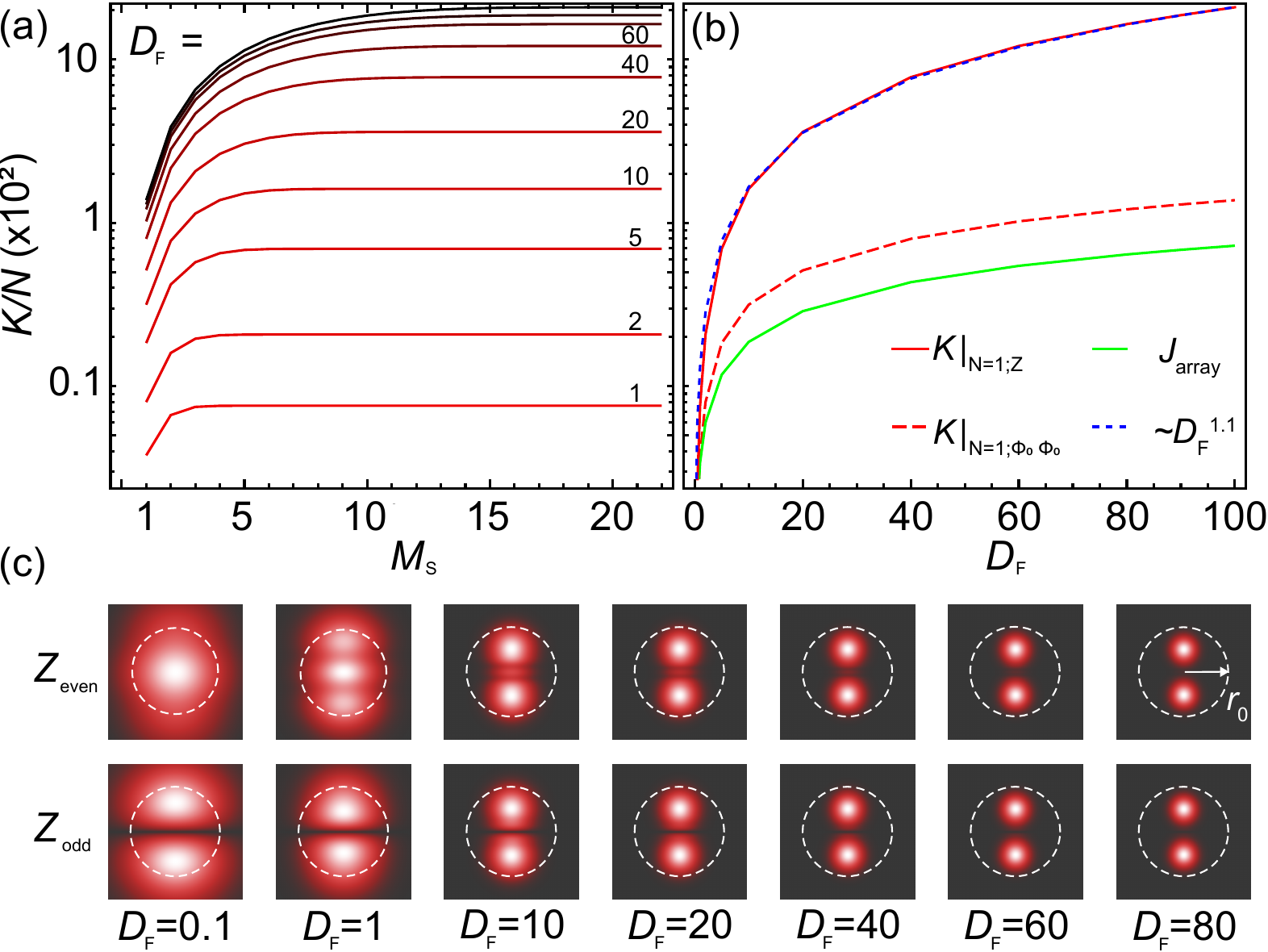}
\caption{(a) Displacement QFI ($K$) versus modal support ($M_\t{s}$) for optical systems with different Fresnel-number product ($D_\text{F}$), normalized to the mean photon number $N$ of the transmitter.  (b) QFI versus $D_\text{F}$ for a transmitter in the fundamental $\t{HG}_{00}$ mode (dashed red) and an optimum spatial mode (solid red). Green is the CFI for the array-based transceiver in Table I. Blue is a power law fit, 
$13.2 D_F^{1.1}$. (c) Even $Z_\t{even}$ ($M_\t{s}\in 2\mathbb{Z}$) and odd $Z_\t{odd}$ optimal modes for different $D_\text{F}$. Any optimal mode can be written as superposition of $Z_\t{even}$ and $Z_\t{odd}$. }
\label{Fig_2}
\end{figure}

The beam is now allowed to propagate through the optical system, whose loss we model  by placing soft Gaussian apertures of radius $r_{T}$ and $r_\t{R}$ at the transmitter and receiver, respectively, separated by a distance $L$.  The net loss is characterized by Fresnel number product $D_\text{F} = (kr_\t{T}r_\t{R}/4L)^2$, which approximates the number of HG modes that can pass through the optical system with negligible loss.  If the beam undergoes a transverse displacement $d\ll r_\t{R}$ during propagation, then the field beyond the receiver aperture can be expressed as
\begin{equation}\label{eq:5}
  \psi(x+d,y)\approx \sqrt{N}\sum\displaylimits^{M_\t{s}-1}_{m,n=0} Q_{mn}(\tilde{d})\sqrt{\eta^{n+1}} c_n e^{i \theta_n} \phi_{m}(x)\phi_0(y),
\end{equation}
where 
\begin{equation}
\eta=\frac{1+2D_\text{F}-\sqrt{1+4D_\text{F}}}{2D_\text{F}}
\end{equation}
is the transmissivity of the $\t{HG}_{00}$ mode, $Q_{mn}(\tilde{d})$ is a cross-talk matrix characterizing spatial mode coupling, and for convenience we define the normalized displacement $\tilde{d}\equiv d/r_\t{R}$~\cite{SI}. 

Eq. \ref{eq:5} implies that for a given transmitter mode, all information about displacement is encoded in the cross-talk matrix $Q_{mn}(\tilde{d})$, and that information in higher order modes must be leveraged against transmission loss $\eta^{n+1}$.  Our main result is to formalize this statement, viz, combining Eqs. 2-6 and the cross-talk matrix derived in the appendix, and noting that
\begin{equation}\begin{aligned}
\langle \hat{q}_m\rangle &=2 \mathrm{Re}\left\{\sqrt{N}\sum_{n=m-1}^{m+1}Q_{mn}(\tilde{d}) \sqrt{\eta^{n+1}} c_n e^{i \theta_n} \right\},\\
\langle \hat{p}_m \rangle & =2 \mathrm{Im}\left\{\sqrt{N}\sum_{n=m-1}^{m+1}Q_{mn}(\tilde{d}) \sqrt{\eta^{n+1}} c_n e^{i \theta_n} \right\},
\end{aligned}\end{equation}
we arrive at the following QFI assuming $r_\t{T} = r_\t{R}\equiv r_0$,
\begin{equation}\begin{aligned}\label{eq:QFIClassical}
   K(\tilde{d})=16\eta N\bigg(\sum\displaylimits_{j=0}^{M_\t{s}-1}\left(j+\left(2j+1\right)D_{\mathrm{f}}\left(1-\eta\right)\right)\eta^{j}c_j^2 -\\
   \eta\sum\displaylimits_{j=0}^{M_\t{s}-3} \sqrt{4\left(j+1\right)\left(j+2\right)D_{\mathrm{f}}} \eta^{j}c_{j} c_{j+2}\sin{\left(\theta_{j}-\theta_{j+2}\right)}\bigg),
\end{aligned}\end{equation}
where $M_\t{s}$ is the highest mode order allowed. 

In Fig. \ref{Fig_2}, we use Eq. \ref{eq:QFIClassical} to visualize the landscape of classically optimal spatial modes for beam displacement sensing, and their performance as a function of the Fresnel number product $D_\text{F}$ of the optical system.  Intuitively, larger $D_\text{F}$ allows for  
a larger spatial mode support $M_\t{s}$, which in turn gives access to higher displacement sensitivity, using an optimal receiver. This reasoning is borne out in Fig. \ref{Fig_2}a and \ref{Fig_2}b, in which the QFI per photon $K/N$ is plotted versus $M_\t{s}$ and $D_\t{F}$, under the constraint $\sum\displaylimits_{n=0}^{M_\t{s}-1}c_\t{n}^2 = 1$.  As shown in Fig. \ref{Fig_2}a, for a fixed $D_\t{F}$, QFI increases with $M_\t{s}$ up to a saturation value of $M_\t{s}^\t{max}\approx \sqrt{2 D_\text{F}}$ \cite{num_of_modes}.  Beyond this value, as visualized in Fig. \ref{Fig_2}b (dashed blue), the maximum QFI scales roughly as $K_\t{max}\approx 13.2 D_\text{F}^{1.1}$ for $D_\text{F}\in(0,100]$, corresponding to a (normalized) displacement imprecision lower bound of 
\begin{equation} \label{eq:qcrb_opt}
    \Delta \tilde{d} \gtrsim \frac{1}{3.6\sqrt{ D_\t{F}^{1.1} N}}.
\end{equation}

The bound in Eq. \ref{eq:qcrb_opt} is the quantum Cram\'{e}r-Rao bound \cite{braunstein1994QFI} (QCRB) for transverse beam displacement sensing through a matched pair of Gaussian apertures.  For comparison,  inserting $(c_0, \theta_0) = (1,0)$ into Eq. \ref{eq:QFIClassical} yields the optimum displacement precision for a conventional $\t{HG}_{00}$ \mbox{coherent state transmitter}
\begin{equation}
    \Delta \tilde{d}_{00} \geq 
    \frac{\sqrt{D_\t{F}/2}}{\sqrt{(\sqrt{1+4 D_\t{F}}-1)^3 N}}\approx \frac{1}{4\sqrt{D_\t{F}^{0.5} N}}.
\end{equation}
Evidently $\Delta \tilde{d}_{00}/\Delta \tilde{d}\approx D_\t{F}^{0.3}$ for $D_\text{F}\in(0,100]$, which implies that a 10-fold reduction in mean squared error is possible, by mode-shape engineering, at $D_\text{F} = 100$.

In order to take advantage of the scaling in Eq. \ref{eq:qcrb_opt}, it is necessary to use an optimal spatial mode shape and receiver. In Fig. \ref{Fig_2}c, we present classically optimal transmitter spatial modes $Z(x,y)$ for various $D_\t{F}$, determined by {maximizing Eq. \ref{eq:QFIClassical} over $\{c_n,\theta_n\}$ for a modal support $M_\t{s}$ high enough to saturate the
QFI, as shown in Fig. \ref{Fig_2}a. 
Evidently $Z$ exhibits a bimodal intensity distribution along the displacement $(x)$ direction, whose root mean square distance from the optical axis increases with the aperture size $r_0$. Not shown is the phase profile of the mode, which exhibits rapid oscillations within the intensity envelope with a period $\sim r_0/\sqrt{D_\t{F}}$.  These oscillations account for the high spatial frequency content of the modeshape, and are practical considerations for receiver architectures at high $D_\t{F}$, as discussed below.

We now turn our attention to the identification of optimal receivers, using CFI 
$(J)$ 
as a figure of merit. 
Formally, for a specific measurement applied to a displaced laser beam
\begin{equation}
J(\tilde{d}) =\int \left(\frac{\partial P(\vec{\gamma};\tilde{d})}{\partial \tilde{d}}\right)^2 \frac{1}{P(\vec{\gamma};\tilde{d})}d\vec{\gamma},
\end{equation}
where $P(\vec{\gamma};\tilde{d})$ is the probability of measurement outcome $\vec{\gamma}$ given normalized displacement $\tilde{d}$.  To identify an optimal receiver for a given transmitter mode, we demand that it satisfy the upper bound $J(\tilde{d}) = K(\tilde{d})$ in Eq. \ref{eq:1}, which corresponds to an optimal displacement imprecision $\Delta \tilde{d} = 1/\surd J(\tilde{d})$ saturating the QCRB given in Eq. \ref{eq:qcrb_opt}. (We note that for single parameter estimation, the existence of such an optimal receiver is assured \cite{braunstein1994QFI}.})


 \begin{figure}[b!]
 \vspace{-1mm}
\centering  \includegraphics[width=0.6\columnwidth]{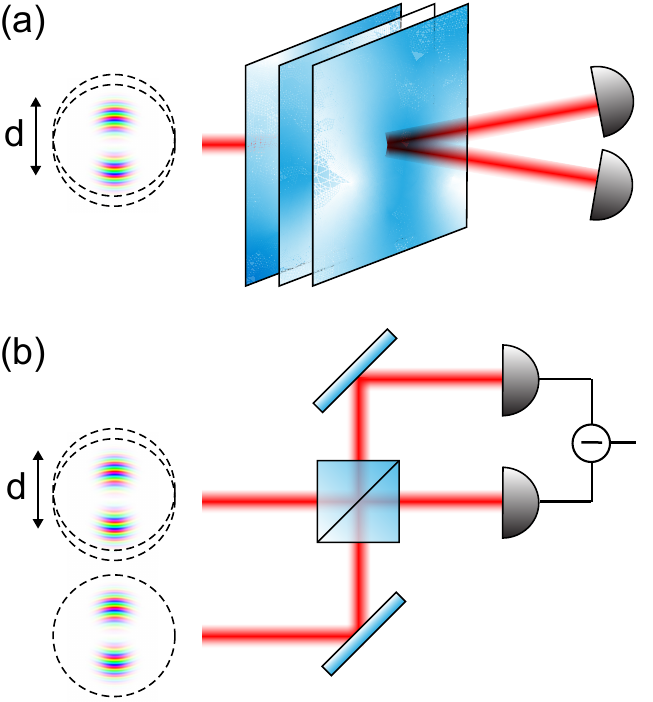}
\caption{Two optimal receivers for transverse beam displacement: (a) a spatial mode sorter which has been configured to distill the optimal transmitter mode $Z$ from its orthogonal complement, and (b) a homodyne interferometer with a local oscillator in the $\zeta_1$ mode. }
\label{Fig_3}
\end{figure}


\begin{table}[b!]
\vspace{-1mm}
\begin{ruledtabular}
\begin{tabular}{ l l l }
Transmitter&Receiver & QFI attaining? \\ 
\colrule
$Z(x)\Phi_0(y)$ & HGSPADE-DD \cite{fontaine2019MODESORTER, zhou2018HG_sorter, spade_displacementsensing2020}& Yes\\ 
$Z(x)\Phi_0(y)$ & $\zeta$-SPADE-DD \cite{ozer2022reconfigurablemodesorter}& Yes \\ 
$Z(x)\Phi_0(y)$ & HGSPADE-Homodyne \cite{2014-QFI-HGn0}& Yes\\
$Z(x)\Phi_0(y)$ & $\zeta$-Homodyne \cite{SI}& Yes \\ 
$\Phi_0(x)\Phi_0(y)$ & HGSPADE-DD \cite{fontaine2019MODESORTER, zhou2018HG_sorter, spade_displacementsensing2020}& Yes \\ 
$\Phi_0(x)\Phi_0(y)$ & ARRAY-DD \cite{TEM00displacement}& No\\ 
\end{tabular}
\caption{Classical transceiver designs and their optimality.}
\end{ruledtabular}
\label{table:1}
\end{table}

We focus on optimal receiver designs for a coherent state transmitter in an odd-ordered optimal spatial mode $Z(x, y)$ (subscript omitted), shown in Fig. \ref{Fig_2}c.  After propagation, it is convenient to express the received field in the form \cite{SI}
\begin{equation}\begin{aligned}\label{eq:rfield_opt}
\psi_\t{opt}
 &= \sqrt{N}\sqrt{\eta_Z}\zeta_0(x+d, y) 
  A_\t{d}(x,y)\\ 
 &\approx \sqrt{N}\bigg(\sum_{n=0}^{M_\t{s}/2-1} \alpha_{2n+1}\eta^{n+1} e^{i n \pi/2} \phi_{2n+1}(x),\\
&\hspace{8mm}+\tilde{d} \sum_{n=0}^{M_\t{s}/2} \beta_{2n} e^{i(\vartheta_{2n}+\vartheta_0)} \phi_{2n}(x)\bigg)\phi_{0}(y),\\
&\equiv\sqrt{N} \sqrt{\eta_Z}\bigg(\zeta_0(x,y)+\sqrt{\boldsymbol{\beta}}\tilde{d}\zeta_1(x,y)\bigg),
\end{aligned}\end{equation}
where $\{\zeta_0,\zeta_1\}$ form an orthonormal principal component basis for the receiver and  $A_\t{d} = e^{(d^2+2x d)/r_R^2}$ is a correction factor for the displaced receiver aperture \cite{SI}. If there is no displacement, photons in transmitter mode $\mathrm{Z}$ will occupy mode $\zeta_0$ at the receiver, with a transmissivity $\eta_Z$. 
Displacement will cause photons to populate mode $\zeta_1$ with coupling strength $\boldsymbol{\beta}$.

For sufficiently small beam displacement $d$, Eq. \ref{eq:rfield_opt} implies that  $d$ is fully encoded in the complex amplitude of mode $\zeta_1$, and that, therefore, the optimal receiver design is independent of the magnitude of $d$. Since $d$ is encoded in a phase space displacement of $\zeta_1$ \cite{weedbrook2012gaussian}, it can be shown that a homodyne receiver with its local oscillator (LO) in the $\zeta_1$ mode is optimal \cite{oh2019optimal}. When only the absolute displacement magnitude $|d|$ is of interest, then phase insensitive mode sorting based receivers are also optimal.  In general, homodyne and SPADE are examples of a broader class of optimal receivers whose defining characteristic is the ability to resolve $\zeta_1$ from $\zeta_0$.

In Figure 3 we illustrate the concept of spatial mode demultiplexing (SPADE) and ``structured-LO''  homodyning, representing two broad classes of optimal coherent state receivers for transverse beam displacement. A more detailed list is given in Table I, together with the traditional, sub-optimal approach based on direct detection with a pixel array (DD-ARRAY).  Below, we elaborate on the CFI analysis of SPADE, structured homodyne, and  DD-ARRAY receivers, and under what circumstances they can be optimal.

For simplicity, we first consider the structured homodyne receiver illustrated in Fig. \ref{Fig_3}b.  By interfering the received field with a LO in the information-carrying $\zeta_1$ mode, an appropriate LO phase yields a (real) quadrature estimate with mean $\langle \hat{q} \rangle = \sqrt{4\eta_Z N\beta}\tilde{d}$ and variance $\langle \Delta\hat{q}\rangle $ = 1~\cite{weedbrook2012gaussian}, \mbox{corresponding to}
\begin{equation}
J_\t{hom}(\tilde{d}) = 4 \eta_Z N\beta.
\end{equation}
Rewriting Eq. \ref{eq:QFIClassical} in the principal component basis $\{\zeta_0,\zeta_1\}$, it is straightforward to show that $J(\tilde{d})=K(\tilde{d})$ \cite{SI}, implying that structured homodyne is indeed an optimal receiver.

For the SPADE receiver (Fig. \ref{Fig_3}a), 
the displaced beam is passed through a series of phase plates that sort principal component  modes $\zeta_0$ and $\zeta_1$ into different paths. (Mode sorting in the HG and other orthonormal bases has been widely realized \cite{fontaine2019MODESORTER, zhou2018HG_sorter, spade_displacementsensing2020}; we here consider a reconfigurable mode sorter \cite{ozer2022reconfigurablemodesorter}.)  Direct photon counting is then carried out in the $\zeta_1$ beam path.  The set of possible counts $\{n_1\}$ follows a Poisson distribution with mean value $N_1 = \eta_Z N \beta \tilde{d}^2$, yielding a CFI
\begin{equation}    \label{eq:spadecfi}
    J_\t{SPADE}(\tilde{d}) = \sum_{n_1=0}^\infty \left(\frac{\partial P(n_1;\tilde{d})}{\partial \tilde{d}}\right)^2 \frac{1}{P} = \left(\frac{\partial N_1}{\partial \tilde{d}}\right)^2 \frac{1}{N_1},
\end{equation}
where $P(n_1;\tilde{d})$ is the conditional probability of $n_1$ counts. It follows from the right hand side of Eq. \ref{eq:spadecfi} that
\begin{equation}
J_\t{SPADE}(\tilde{d}) = 4 \eta_Z N\beta,
\end{equation}
implying that SPADE in the $\{\zeta_0,\zeta_1\}$ basis, together with direct photon counting of the $\zeta_1$-sorted light, is also an optimal receiver. In Ref. \cite{SI}, we show that SPADE in an arbitrary orthonormal basis, followed by direct photon counting or homodyne of each output, is also optimal (HGSPADE-DD and HGSPADE-homodyne in Table I, respectively).

 \begin{figure}[b!]
 \vspace{-2mm}
\centering  
\includegraphics[width=0.9\columnwidth]{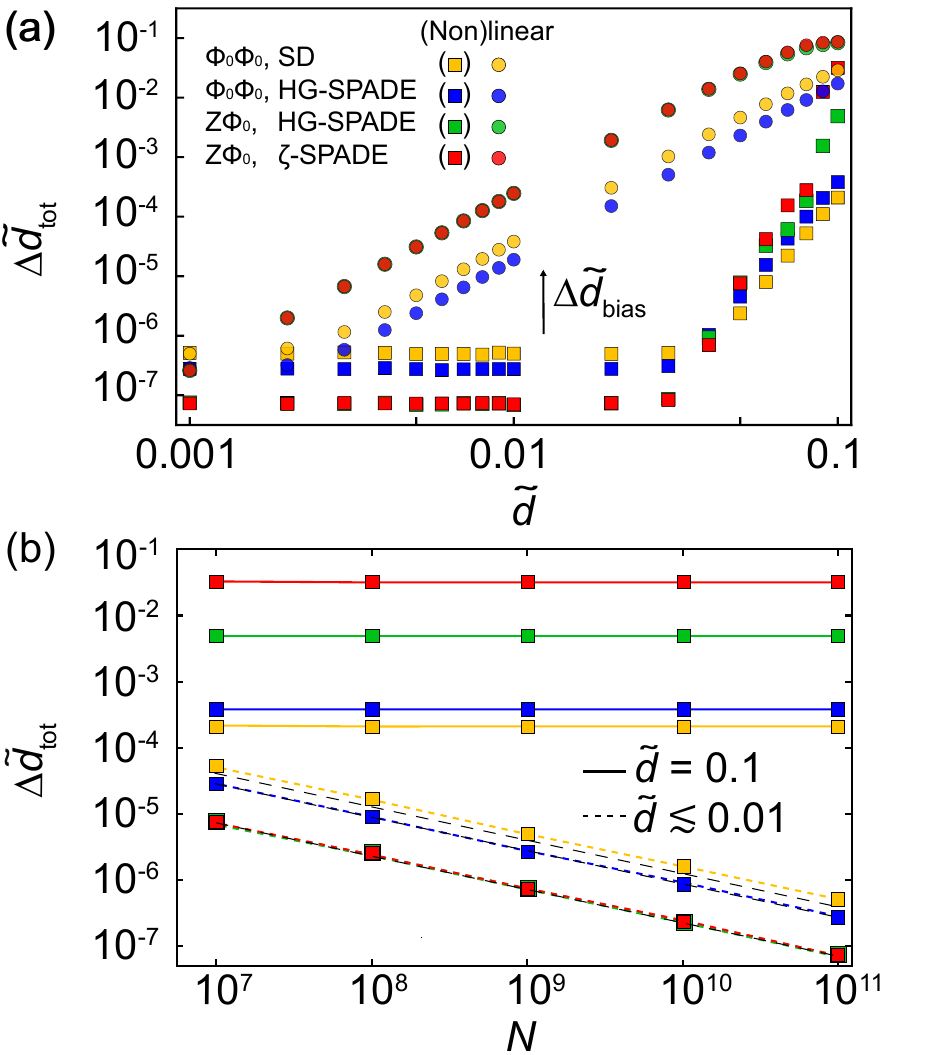}
\caption{(a) Simulated total measurement error versus actual displacement for various transceiver architectures, using linear and nonlinear estimators. (b) Simulated total measurement error versus photon number in the shot noise (dashed) and bias (solid) limited regime. }
\label{Fig_4}
\vspace{-2mm}
\end{figure}

Finally, as an example of a non-ideal receiver, we consider a traditional pixel array detector together with photon counting at each pixel.  
Pixel array detectors have been studied extensively for transverse beam displacement sensing with both classical and squeezed light \cite{2000-SD-theory,2002-SD-EXP1D,2003-SD-EXP2D, LEPD2022, TEM00displacement,camera_fisherinformation} 
(including split photodetection, which corresponds to a $1\times2$ pixel array \cite{2000-SD-theory,2002-SD-EXP1D,2003-SD-EXP2D, LEPD2022}). Following the typical case, we assume the transmitter is a coherent state in the HG$_{00}$ mode. For simplicity, we also assume that the pixels are infinitesimally small with 100\% fill factor~\cite{TEM00displacement, camera_fisherinformation}. The output of this receiver is an infinite set of Poisson distributed random variables with mean $\sigma_N(x,y,\tilde{d}) \t{d}x\t{d}y$, where 
\begin{equation}
   \sigma_N(x,y,\tilde{d}) = N\eta|\phi_0(x+\tilde{d} r_R)\phi_0(y)A_\t{d}(x,y)|^2
\end{equation}
is the mean count density.  Generalizing Eq. \ref{eq:spadecfi} then yields
\begin{equation}
J_\t{array}(\tilde{d})=\iint\left(\frac{\partial \sigma_N}{\partial \tilde{d}}\right)^2\frac{1}{\sigma_N}\t{d}x\t{d}y=4 \eta N  \frac{(\sqrt{1+4D_\text{F}}-1)^2}{\sqrt{1+4D_\text{F}}}, 
    \label{CFI_F}
\end{equation}
where the integral is over all space.


We conclude by discussing the dynamic range of the receivers in Table I, particularly those employing an optimal transmitter mode $Z$ with large modal support $M_\t{s}\gg 1$, whose characteristic length scale $\sim r_0/M_\t{s}$ can be much smaller than the receiver aperture (c.f. Fig. \ref{Fig_3}).  In this case, the small displacement approximation $\tilde{d} \ll 1$ underpinning Eq. \ref{eq:rfield_opt} may  be inappropriate, manifesting as a receiver-specific bias error $\Delta \tilde{d}_\t{bias}$ that increases the total measurement error to
\begin{equation}
\Delta \tilde{d}_\t{tot} = \sqrt{\Delta \tilde{d}^2 +\Delta \tilde{d}_\t{bias}^2}.
\end{equation}

In Fig. \ref{Fig_4}, we present Monte Carlo simulations of total error $\Delta \tilde{d}_\t{tot}$ for transceivers employing pixel array and SPADE receivers with $\t{HG}_{00}$ and $Z$-mode coherent state transmitters. (Homodyne and SPADE suffer similar dynamic range limitations; we focus on SPADE because of its simpler extension to different mode bases.)  Specifically, we compare the canonical two-pixel (split photodiode) receiver to SPADE receivers employing an HG mode sorter \cite{spade_displacementsensing2020} 
and a reconfigurable mode sorter in the $\{\zeta_0,\zeta_1\}$ basis, for a system with $D_\t{F}=90$. Based on the standard maximum likelihood estimator, we devise a linear and nonlinear estimator for processing the receiver output (see \cite{SI}).  As expected, we find that $Z$-mode transceivers have smaller imprecision $\Delta \tilde{d}$ and higher bias error $\Delta \tilde{d}_\t{bias}$ than HG$_{00}$ transceivers; however, the bias error can be reduced using a nonlinear estimator (at the expense of computational overhead) yielding a total error that approaches the QCRB for $\tilde{d}\le 0.01$, indicated by the black dashed lines in Fig.\ref{Fig_4}b.
We emphasize that the dynamic range---the displacement $\tilde{d}$ at which $\Delta \tilde{d}_\t{bias} =  \Delta\tilde{d}$---depends on measurement strength $N$, and can in principle be extended---for both linear and nonlinear estimators---by active stabilization of the beam position.

In summary, we have used quantum Fisher information (QFI) to design an optimal transceiver for laser beam displacement sensing.  Assuming the probe is in a coherent state and modeling propagation loss as a pair of Gaussian apertures at the transmitter (laser) and receiver, we find that the optimal spatial mode is a bimodal distribution (Fig. \ref{Fig_2}) that balances the trade-off between maximizing spatial frequency content and minimizing aperture loss.  We emphasize that this spatial mode is optimized over all degrees of freedom of an optical system, including possible receivers, and is parameterized only by the system's Fresnel number product $D_\t{F}$.  Its mean-squared error relative to the traditional fundamental Gaussian mode transceiver scales as $\sim D_\t{F}^{0.6}$, yielding a 10-fold improvement for practically relevant $D_\t{F}\sim 100$.  We also studied various receiver architectures, using classical Fisher information (CFI) as a metric, and showed that homodyne and SPADE receivers can each extract maximal information (CFI = QFI) about beam displacement when appropriately tailored to the transmitter mode (Fig. \ref{Fig_3}), while traditional pixel array receivers cannot.  Finally, we considered the dynamic range of SPADE and 2-pixel (split detector) receivers, and showed that the increased bias error inherent to the optimal spatial mode can in principle be reduced with a nonlinear estimator (Fig.~\ref{Fig_4}).

Looking forward, we emphasize that our results are both practically relevant, given recent rapid advances in structured light preparation \cite{SLM-structured-light}, and extensible to non-classical probe states, such as squeezed light. Combining these two resources---spatial mode structuring and squeezing---can provide access to Heisenberg scaling $\Delta d\sim 1/N$ \cite{quantummetrology,2023mode_parameter_estimation} for beam displacement measurements and related imaging tasks (such as deflectometry), and has close connections to recent work in entanglement-enhanced distributed sensing \cite{xia2023entanglement,grace2020entanglement}, albeit with a fundamentally different set of challenges and applications related to the spatial mode entanglement encoding.

\section*{Acknowledgements}
WH and SG acknowledge Office of Naval Research (ONR) contract number N00014-19-1-2189, and Air Force Office of Scientific Research (AFOSR) contract number FA9550-22-1-0180 for sponsoring this research. CNG acknowledges funding support from the National Science Foundation, FET, Award No. 2122337. DJW acknowledges support from the National Science Foundation through award number 2239735.


\begin{thebibliography}{35}
\bibitem [{num()}]{num_of_modes}%
  \BibitemOpen
  \href@noop {} {}\bibinfo {note} {The total number of HG modes with near unity transmissivity is approximately $D_\t{F}$. For the 1D problem at hand we have approximately $\sqrt{2 D_\t{F}}$ HG modes with near unity transmissivity when keeping $\Phi_0(y)$ as the mode function along y axis.}\BibitemShut {Stop}%
\end{thebibliography}
%

\newpage

\foreach \x in {1,...,5}
{\clearpage
\includepdf[pages={\x,{}}]{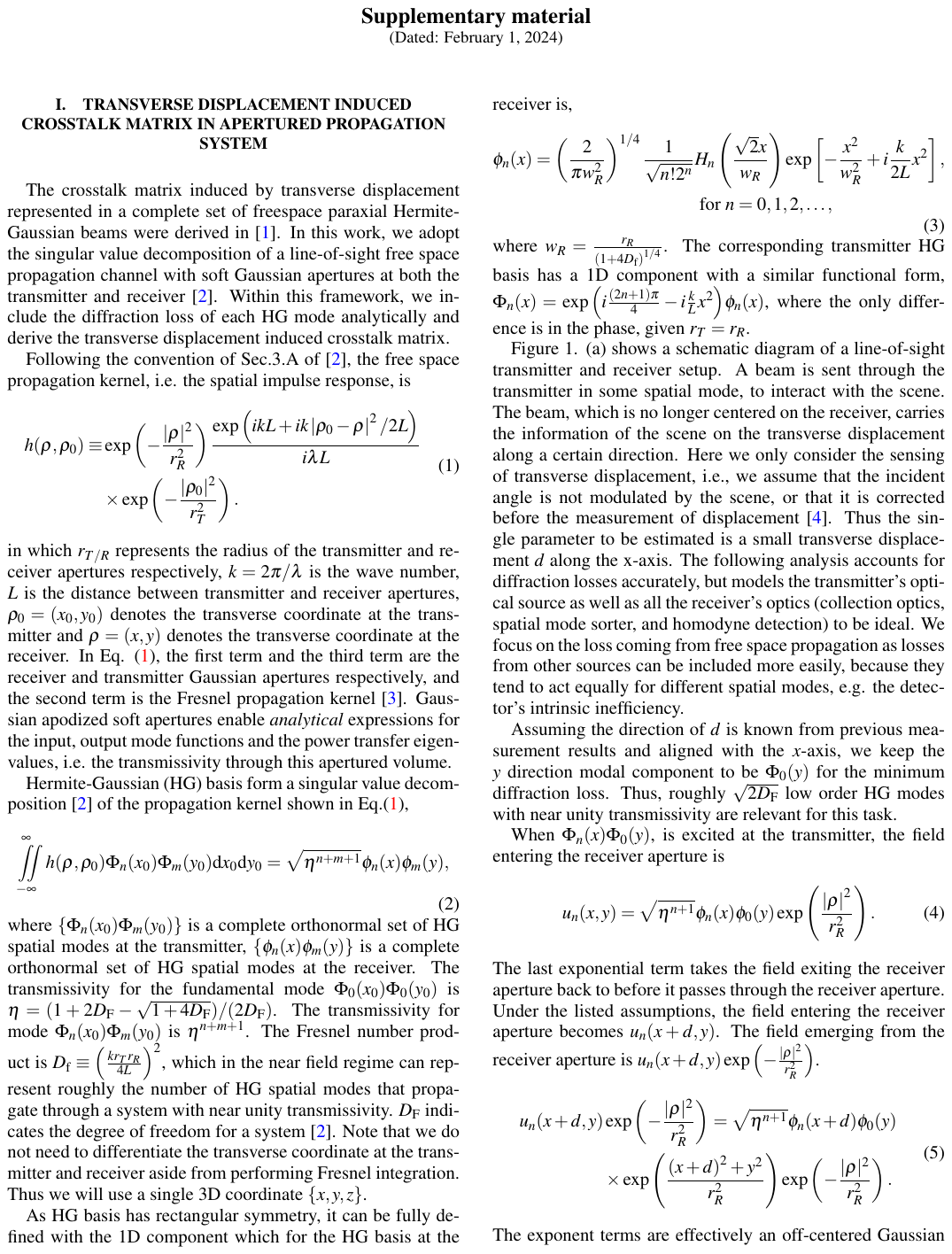}}

\end{document}


\title{Supplementary material}
\date{\today}
\maketitle
\section{Transverse displacement induced Crosstalk matrix in apertured propagation system}
The crosstalk matrix induced by transverse displacement represented in a complete set of freespace paraxial Hermite-Gaussian beams were derived in \cite{qi2018ultimate}. In this work, we adopt the singular value decomposition of a line-of-sight free space propagation channel with soft Gaussian apertures at both the transmitter and receiver \cite{shapiro2005ultimate}. Within this framework, we include the diffraction loss of each HG mode analytically and derive the transverse displacement induced crosstalk matrix. 


Following the convention of Sec.3.A of \cite{shapiro2005ultimate},
the free space propagation kernel, i.e. the spatial impulse response, is 
\begin{equation}
\begin{aligned}
h(\boldsymbol{\rho},\boldsymbol{\rho_0})\equiv &\exp \left(-\frac{|\boldsymbol{\rho}|^{2}}{ r_{R}^{2}}\right) \frac{\exp \left(i k L+i k\left|\boldsymbol{\rho_0}-\boldsymbol{\rho}\right|^{2} / 2 L\right)}{i \lambda L} \\
&\times\exp \left(-\frac{|\boldsymbol{\rho_0}|^{2} }{r_{T}^{2}}\right).
\label{kernel}
\end{aligned}
\end{equation}
in which $ r_{T/R}$ represents the radius of the transmitter and receiver apertures respectively, $k=2\pi/\lambda$ is the wave number, $L$ is the distance between transmitter and receiver apertures, $\boldsymbol{\rho_{0}}=\left(x_{0},y_{0}\right)$ denotes the transverse coordinate at the transmitter and $\boldsymbol{\rho}=\left(x,y\right)$ denotes the transverse coordinate at the receiver. In Eq. \eqref{kernel}, the first term and the third term are the receiver and transmitter Gaussian apertures respectively, and the second term is the Fresnel propagation kernel \cite{goodman}. Gaussian apodized soft apertures enable \textit{analytical} expressions for the input, output mode functions and the power transfer eigenvalues, i.e. the transmissivity through this apertured volume.

Hermite-Gaussian (HG) basis form a singular value decomposition \cite{shapiro2005ultimate} of the propagation kernel shown in Eq.\eqref{kernel}, 
\begin{equation}
\iint\displaylimits^\infty_{-\infty}h(\boldsymbol{\rho},\boldsymbol{\rho_0}) \Phi_{n}(x_0) \Phi_{m}(y_0) \mathrm{d}x_0\mathrm{d}y_0
=  \sqrt{\eta^{n+m+1}}  \phi_{n}(x)\phi_{m}(y),
\end{equation}
where $\{\Phi_{n}(x_0) \Phi_{m}(y_0)\}$ is a complete orthonormal set of HG spatial modes at the transmitter, $\{\phi_{n}(x)\phi_{m}(y)\}$ is a complete orthonormal set of HG spatial modes at the receiver. The transmissivity for the fundamental mode $\Phi_{0}(x_0)\Phi_{0}(y_0)$ is $\eta=(1+2D_\text{F}-\sqrt{1+4D_\text{F}})/(2D_\text{F})$. The transmissivity for mode $\Phi_{n}(x_0)\Phi_{m}(y_0)$ is $\eta^{n+m+1}$. The Fresnel number product is $D_{\mathrm{f}}\equiv\left(\frac{k r_{T} r_{R}}{4L}\right)^2$, which in the near field regime can represent roughly the number of HG spatial modes that propagate through a system with near unity transmissivity. $D_\text{F}$ indicates the degree of freedom for a system \cite{shapiro2005ultimate}. Note that we do not need to differentiate the transverse coordinate at the transmitter and receiver aside from performing Fresnel integration. Thus we will use a single 3D coordinate $\{x,y,z\}$. 

As HG basis has rectangular symmetry, it can be fully defined with the 1D component which for the HG basis at the receiver is,
\begin{equation}
\begin{aligned}
    \phi_{n}(x)=\left(\frac{2}{\pi w_{R}^{2}}\right)^{1 / 4} \frac{1}{\sqrt{n ! 2^{n}}} &H_{n}\left(\frac{\sqrt{2} x}{w_{R}}\right) \exp \left[-\frac{x^2}{w_{R}^2}+i\frac{k}{2L}x^2\right], \\
    &\text { for } n=0,1,2, \ldots,
    \label{HGmode}    
\end{aligned}
\end{equation}
where $w_R=\frac{r_{R}}{\left(1+4D_{\mathrm{f}}\right)^{1/4}}$. The corresponding transmitter HG basis has a 1D component with a similar functional form, $\Phi_{n}(x)=\exp{\left(i \frac{(2n+1)\pi}{4}-i\frac{k}{L}x^2\right)} \phi_{n}(x)$, where the only difference is in the phase, given $r_T=r_R$.

Figure 1. (a) shows a schematic diagram of a line-of-sight transmitter and receiver setup. A beam is sent through the transmitter in some spatial mode, to interact with the scene. The beam, which is no longer centered on the receiver, carries the information of the scene on the transverse displacement along a certain direction. Here we only consider the sensing of transverse displacement, i.e., we assume that the incident angle is not modulated by the scene, or that it is corrected before the measurement of displacement \cite{grace2020josaa}. Thus the single parameter to be estimated is a small transverse displacement $d$ along the x-axis. The following analysis accounts for diffraction losses accurately, but models the transmitter's optical source as well as all the receiver's optics (collection optics, spatial mode sorter, and homodyne detection) to be ideal. We focus on the loss coming from free space propagation as losses from other sources can be included more easily, because they tend to act equally for different spatial modes, e.g. the detector's intrinsic inefficiency. 

Assuming the direction of $d$ is known from previous measurement results and aligned with the $x$-axis, we keep the $y$ direction modal component to be $\Phi_0(y)$ for the minimum diffraction loss. Thus, roughly $\sqrt{2D_\text{F}}$ low order HG modes with near unity transmissivity are relevant for this task.

When $\Phi_{n}(x) \Phi_{0}(y)$, 
is excited at the transmitter, the field entering the receiver aperture is
\begin{equation}
    u_{n}(x,y)=\sqrt{\eta^{n+1}} \phi_{n}(x) \phi_{0}(y)  \exp \left( \frac{\left|\boldsymbol{\rho}\right|^{2}}{ r_{R}^{2}}\right).
    \label{field_before_rR}
\end{equation}
The last exponential term takes the field exiting the receiver aperture back to before it passes through the receiver aperture. 
Under the listed assumptions, the field entering the receiver aperture becomes
$u_n(x+d,y)$. The field emerging from the receiver aperture is $u_n(x+d,y)\exp \left(-\frac{|\boldsymbol{\rho}|^{2}}{ r_{R}^{2}}\right)$. 
\begin{equation}    
\begin{aligned}
u_{n}(x+d,y)&\exp \left(-\frac{|\boldsymbol{\rho}|^{2}}{ r_{R}^{2}}\right)=\sqrt{\eta^{n+1}} \phi_{n}(x+d) \phi_{0}(y)  \\
&\times \exp \left( \frac{\left(x+d\right)^2+y^2}{ r_{R}^{2}}\right)\exp \left(-\frac{|\boldsymbol{\rho}|^{2}}{ r_{R}^{2}}\right).
\end{aligned}
\end{equation}
The exponent terms are effectively an off-centered Gaussian amplitude aperture,
\begin{equation}
    A_\t{d}(x, y)\equiv \exp \left( \frac{\left(x+d\right)^2+y^2}{ r_{R}^{2}}\right)\exp \left(-\frac{|\boldsymbol{\rho}|^{2}}{ r_{R}^{2}}\right).
\end{equation}
The transverse beam displacement induced crosstalk can be obtained by projecting the emerged field onto the receiver HG modes $\phi_{m}\left(x\right)\phi_{0}\left(y\right)$. 
\begin{widetext}
\begin{equation}
\begin{aligned}
&\iint\displaylimits_{-\infty}^{\infty}\phi_{m}^{\ast}\left(x\right)\phi_{0}^{\ast}\left(y\right) u_n(x+d,y)\exp \left(-\frac{|\boldsymbol{\rho}|^{2}}{ r_{R}^{2}}\right) \mathrm{d}x\mathrm{d}y,\\
     =& \int_{-\infty}^{\infty}\phi_{m}^{\ast}\left(x\right) \sqrt{\eta^{n+1}} \phi_{n}\left(x+d\right) \exp \left( \frac{(x+d)^{ 2}}{ r_{R}^{2}}\right) \exp\left(-\frac{x^2}{r_{R}^{2}}\right) \mathrm{d}x,\\
     =& \sqrt{\eta^{n+1}} \exp \left(\frac{d^2}{r_{R}^2}\right) \int_{-\infty}^{\infty}\phi_{m}^{\ast}\left(x\right)  \phi_{n}\left(x+d\right) \exp \left( \frac{2 x d}{ r_{R}^{2}}\right)\mathrm{d}x,\\
     =& \sqrt{\eta^{n+1}} \exp \left(\frac{d^2}{r_{R}^2}\right) \int_{-\infty}^{\infty} \phi_{m}^{\ast}\left(x\right)\left[\phi_{n}\left(x\right)+d\phi_{n}^{\prime}\left(x\right)+\mathcal{O}(d^2) \right]\left[1+\frac{2 x d}{r_R^2} +\mathcal{O}(d^2)\right]\mathrm{d}x,\\
  =&\sqrt{\eta^{n+1}} \left( \delta_{m,n}+\frac{d w_{R}}{r_{R}^2}\left(\sqrt{m+1} \delta_{m+1,n}+\sqrt{m} \delta_{m-1,n}\right)+d\left(-\frac{1}{w_R}+\frac{i  k w_{R}}{2L}\right)\sqrt{m} \delta_{m-1,n}+d\left(\frac{1}{w_R}+\frac{i  k w_{R}}{2L}\right)\sqrt{m+1} \delta_{m+1,n}+\mathcal{O}(d^2)\right),\\
 \approx &\sqrt{\eta^{n+1}} \left( \delta_{m,n}+\frac{d}{w_R}\left(\sqrt{m+1} \delta_{m+1,n}-\sqrt{m} \delta_{m-1,n}\right) +\frac{d w_{R}}{r_{R}^2}\left(\sqrt{m+1} \delta_{m+1,n}+\sqrt{m} \delta_{m-1,n}\right)+\frac{i d k w_{R}}{2L}\left(\sqrt{m+1} \delta_{m+1,n}+\sqrt{m} \delta_{m-1,n}\right)\right),
\label{lineardp}
\end{aligned}
\end{equation}
\end{widetext}

To calculate the modal crosstalk, the recurrence relations for Hermite polynomial are used to deduce the recurrence relation of $\phi_{n}(x)$. 
\begin{equation}\footnotesize
\phi_n(x)^{\prime}= \frac{w_{R}}{2   }(i\frac{k}{L}-\frac{2}{w_{R}^2}) \sqrt{n+1}\phi_{n+1}(x)+(i\frac{kw_R}{2L}+\frac{1}{w_{R}})\sqrt{n} \phi_{n-1} (x),
\end{equation}
\begin{equation}\footnotesize
      \phi_{n+1}(x)=\frac{2  x }{w_{R}\sqrt{n+1}} \phi_{n}(x)-\sqrt{\frac{n}{n+1}} \phi_{n-1} (x),
\end{equation}

The last line of Eq. \eqref{lineardp} shows that in the limit of small $d$, the diffraction loss is the same, $\eta^{n+1}$. The transverse displacement induced crosstalk matrix is,

\begin{equation}\footnotesize
\begin{aligned}
Q(\tilde{d})\equiv &\boldsymbol{\mathcal{I}}+\frac{d}{w_{R}} \boldsymbol{\Gamma}+\left(\frac{d w_{R}}{r_{R}^2}+\frac{i d k w_{R}}{2L}\right) |\boldsymbol{\Gamma}|,\\
=&\boldsymbol{\mathcal{I}}+\tilde{d}\left[(1+4D_\text{F})^{1/4} \boldsymbol{\Gamma}+\frac{ |\boldsymbol{\Gamma}|}{(1+4D_\text{F})^{1/4}}+\frac{i r_{R} 2 \sqrt{D_\text{F}}}{r_{T} (1+4D_\text{F})^{1/4}} |\boldsymbol{\Gamma}|\right]\\
=&\boldsymbol{\mathcal{I}}+\tilde{d} Q^\prime. \\
\end{aligned}
\label{strength}
\end{equation}
in which $\tilde{d}=d/r_R$, is the normalised transverse displacement w.r.t the receiver aperture radius. $\boldsymbol{\mathcal{I}}$ is an identity matrix and
$\boldsymbol{\Gamma}$ of the first $M_\t{s}$ modes is
    \begin{equation}
    \boldsymbol{\Gamma}(M_\t{s})=\left[\begin{array}{ccccc}
0 & 1 & 0 & \ldots & 0 \\
-1 & 0 & \sqrt{2} & \ldots & 0 \\
0 & -\sqrt{2} & 0 & \ddots & \vdots \\
\vdots & \vdots & \ddots & \ddots & \sqrt{M_\t{s}-1} \\
0 & 0 & \ldots & -\sqrt{M_\t{s}-1} & 0
\end{array}\right]
\end{equation}

The crosstalk matrix in the small transverse displacement limit, is a tri-diagonal matrix. The absolute square of the elements from the crosstalk matrix represents fractional power transfer induced by $d$ along the $x$-axis from $\phi_{n}(x)\phi_0(y)$ to $\{\phi_{n-1}(x)\phi_0(y),\phi_{n}(x)\phi_0(y),\phi_{n+1}(x)\phi_0(y)\}$. We will use the crosstalk matrix $Q(\tilde{d})$ to find the optimal spatial mode to excite a coherent state in, at the transmitter. 

Gaussian apodized apertures provide a physical constraint for a system and also suppress edge diffraction from hard apertures \cite{siegman1986lasers}, which are also explored in many areas where apodization is useful, e.g. coronagraphy \cite{goodman}.
When we have a symmetrical setup, $r_T = r_R \equiv r_0$, with a big $D_\text{F}$, the beam focus is at $z = L/2$. The transmitter and receiver are effectively located at the Rayleigh length away from the waist. The Guoy phase shift accumulated by passing through the focus is effectively shown in the transmitter mode set. Having the beam focus in the middle can be beneficial. Atomic force microscopes are one example, in which the laser beam reflects off the back of the cantilever into the receiver. Setting the beam waist at the middle makes sure the beam is simply deflected after interacting with the scene \cite{putman1992afm}.

\section{Appendix B: Classical Fisher information of $Z$ mode transmitter}
In this section, we present the QFI analysis in the principal component (PC) basis for $Z$-mode coherent state transmitter and the CFI analysis for QFI achieving receivers. Specifically, we present the CFI analysis for HG-SPADE and $\zeta$-SPADE.

$Z$-mode is the spatial mode found by applying Lagrange multiplier method to Eq. 8 in the main text. Therefore the QFI with a $Z$-mode coherent state transmitter is,
\begin{equation}
    K_{Z}\approx 13.2 D_\text{F}^{1.1} N,
\end{equation}
where $N$ is the total mean photon number of the coherent state sent at the transmitter. Representing the odd $Z$-mode in the transmitter side HG basis, we have,
\begin{equation}
Z_\text{odd}(x,y)\equiv\sum_{n=0}^{M_\t{s}/2-1} \alpha_{2n+1} e^{\frac{i n \pi}{2}} \Phi_{2n+1}(x)\Phi_{0}(y),
\end{equation}
in which the coefficients $\left(\alpha_{2n+1},\theta_{2n+1}=n\pi/2\right)$ are results of the optimization. We will omit the subscript \textbf{odd}, for notation simplicity. After propagation through the system, the corresponding receiver mode is,
\begin{equation}
\zeta_0(x,y)\equiv\frac{1}{\sqrt{\eta_Z}}\sum_{n=0}^{M_\t{s}/2-1} \alpha_{2n+1} e^{\frac{i n \pi}{2}} \eta^{n+1} \phi_{2n+1}(x)\phi_{0}(y),
\end{equation}
in which $\eta_Z=\sum_{n=0}^{M_\t{s}/2-1} \alpha_{2n+1}^2  \eta^{2n+2} $ is the transmissivity for mode $Z$. Apply the crosstalk matrix we can represent the parameter carrying field in the HG basis,
\begin{equation}
\begin{aligned}
\psi_\t{opt}(x+d,y)
 &\approx \sqrt{N}\bigg(\sum_{n=0}^{M_\t{s}/2-1} \alpha_{2n+1}\eta^{n+1} e^{i n \pi/2} \phi_{2n+1}(x),\\
&\hspace{8mm}+\tilde{d} \sum_{n=0}^{M_\t{s}/2} \beta_{2n} e^{i(\vartheta_{2n}+\vartheta_0)} \phi_{2n}(x)\bigg)\phi_{0}(y).\\
\end{aligned}
\end{equation}
Note that due to the tri-diagonal form of the crosstalk matrix, the parameter $\tilde{d}$ is encoded on the amplitude of many even HG modes. Now we identify the PC basis. We can alternatively represent the parameter carrying field in the form,
\begin{equation}
\begin{aligned}
\psi_\t{opt}(x+d,y)
&\approx\sqrt{N} \sqrt{\eta_Z}\bigg(\zeta_0(x,y)+\sqrt{\boldsymbol{\beta}}\tilde{d}\zeta_1(x,y)\bigg).
\end{aligned}
\end{equation}
Such that the information-carrying mode,
\begin{equation}
\zeta_1(x,y)\equiv\frac{1}{\sqrt{\beta \eta_Z}} \sum_{n=0}^{M_\t{s}/2} \beta_{2n} e^{i(\vartheta_{2n}+\vartheta_0)} \phi_{2n}(x)\phi_{0}(y),
\end{equation}
where $\beta=\sum_{n=0}^{M_\t{s}/2} \beta_{2n}^2/\eta_Z$ represents the coupling strength from $\zeta_0$ to $\zeta_1$ induced by a small transverse displacement. Up to this point, we identify the PC basis to guide the receiver design, $\{\zeta_0,\zeta_1\}$.

\textbf{QFI analysis in the PC basis for $Z$ mode coherent state transmitter} follows the same procedure given in the main text. Using the mean vector representation of a coherent state, the parameter carrying field can be represented as a two-mode state, $|2\sqrt{N \eta_Z}\rangle_{\zeta_0}|2\sqrt{N \eta_Z\beta} \tilde{d}\rangle_{\zeta_1}$. Applying Eq. 2 and Eq. 3 from the main text we have,
\begin{equation}
    K_Z=4 N \eta_Z \beta. 
\end{equation}
This result further strengthens our conclusion that the total amount of information is in a delicate balance between the diffraction loss and transverse displacement induced coupling strength. For $Z$ mode, we have $4 \eta_Z \beta \approx 13.2 D_\t{F}^{1.1}$. 

\textbf{HG-SPADE receiver} will produce a series of independent random variables each following a Poisson distribution with mean $^\t{HG}N_\t{i}$. For the particular case of $Z_\t{odd}$ mode illumination, the parameter is encoded on the even HG modes' mean photon number,
\begin{equation}
    ^\t{HG}N_\t{i=2n}=N \beta_{2n}^2\tilde{d}^2.
\end{equation}
The corresponding CFI is,
\begin{equation}
    J_\t{HG-SPADE}=\sum_{n=0}^{M_\t{s}/2} \left(\frac{\partial  ^\t{HG}N_\t{2n}}{\partial \tilde{d}}\right)^2\frac{1}{ ^\t{HG}N_\t{2n}}=4 N \sum_{n=0}^{M_\t{s}/2} \beta_{2n}^2.
\end{equation}
Thus it is obvious that HG-SPADE is a QFI saturating receiver. For a system with $D_\t{F}=90$, we need a HG-SPADE receiver that can handle at least 8 even HG modes.

\textbf{$\zeta$-SPADE receiver} on the other hand only sorts two modes. Which in principle should be less demanding than a general HG-SPADE.
A schematic diagram of a spatial mode sorter is given in Fig. 3(a) of the main text. Tow bucket detectors are used to measure light in $\zeta_0$ and $\zeta_1$ respectively. As shown in the main text, the CFI is,
\begin{equation}
   J_\t{\zeta-SPADE}= J_\t{SPADE}=4 N \eta_Z \beta.
\end{equation}
It is straightforward to extend the SPADE-based receiver CFI analysis to homodyne receiver analysis. We claim without proof that HG-SPADE-homodyne is also QFI achieving. 
\section{Monte Carlo simulation for dynamic range analysis}

In this section, we present the linear estimator and the nonlinear estimator (NL) devised from the MLE to demonstrate the classical optimal transceiver design can outperform the conventional transceiver design over some dynamic range by the factor predicted by the Fisher information analysis.

The optimal classical transceiver design based on QFI and CFI analysis is valid in the limit of small transverse displacement. The inverse of them provide a tight lower bound for the mean squared error of unbiased estimators. Maximum likelihood estimator (MLE) in the large trial limit is unbiased and can asymptotically saturate this limit \cite{braunstein1992MLE}.
\begin{equation} 
\mathrm{min}_{\hat{\rho}} \frac{1}{ K(\tilde{d})}  \leq \frac{1}{J(\tilde{d})}
\leq \mathrm{MSE}(\hat{d}_{ML}),
\end{equation}
in which $\mathrm{min}_{\hat{\rho}}$ represents a minimization performed with respect to the transmitted state. We obtained one version of the above QFI and CFI in the limit of very small transverse displacement. To demonstrate the practicality of the optimal transceiver design, we perform Monte Carlo simulations for four transceiver designs, with $\tilde{d}$ sampled discretely within $\left(0,0.1\right)$. We process the measurement results with estimators devised from MLE. The total estimation error $\Delta \tilde{d}_\t{tot}$ approximately saturates the CRB correspondingly for each transceiver for $\tilde{d}\le 0.01$.  



We first redefine the four transceivers under study. Then we review the ML estimation method, and define two estimators by approximating the ML estimator, namely a linearized estimator and a NL estimator. Then the simulated total estimation error is presented for different estimators, different $\tilde{d}$ values, and different total mean photon numbers consumed specifically for a system with $D_\text{F}=90$.

\subsection{Four transceiver designs}
Here, we adapt the transceivers' functionality to incoporate some modifications for a practically bigger dynamic range.
\begin{itemize}
    \item $Z$-mode coherent state transmitter and HG-SPADE: A coherent state is sent through the transmitter in the spatial mode $Z(x, y)$. The receiver will perform mode sorting for HG modes, $\{\phi_j(x)\phi_0(y)\}$, followed by photon counting in each spatial mode. 
    \item $Z$-mode coherent state transmitter and $\zeta$-SPADE: A coherent state is sent through the transmitter in the spatial mode $Z(x, y)$. The receiver will perform mode sorting in $\{\zeta_0(x,y),\zeta_1(x,y)\}$ mode set, followed by photon counting in the two spatial modes instead of just performing photon counting in $\zeta_1(x, y)$.
    \item $\t{HG}_{00}$-mode coherent state transmitter and HG-SPADE: A coherent state is sent through the transmitter in the spatial mode $\Phi_0(x)\Phi_0(y)$. The receiver will perform mode sorting for HG modes, $\{\phi_j(x)\phi_0(y)\}$, followed by photon counting in each spatial mode.
    \item $\t{HG}_{00}$-mode coherent state transmitter and split photo detector: A coherent state is sent through the transmitter in the spatial mode $\Phi_0(x)\Phi_0(y)$. The receiver will perform photon counting with two pixels side-by-side along the $x$-axis, i.e. a split photo detector.
\end{itemize}
The first and the third transceivers listed above are the same as in the small transverse displacement limit. The second transceiver 
sorts two modes instead of one. 
For the fourth transceiver design, we simulate a conventional split photo detector \cite{2000-SD-theory,2002-SD-EXP1D,2003-SD-EXP2D}. All these receivers will produce measurement results following Poisson distribution as we constrain ourselves with coherent state and assume ideal measurement. 

\subsection{Estimator based on Maximum likelihood estimation}
When a set of measurement results is available, the maximum likelihood (ML) estimator perform a maximization on the log-likelihood function with respect to the parameter to be estimated, $\tilde{d}$, for this given set of measurement outcome, 
\begin{equation}
  \hat{d}_{ML}(\vec{\gamma}_\mathrm{meas})=  \mathrm{Maxarg}  \log {P\left( \vec{\gamma}=\vec{\gamma}_\mathrm{meas} ; \tilde{d}\right)},
    \label{pdf}
\end{equation}
in which $\vec{\gamma}_\mathrm{meas}$ represents one set of measurement outcome. One can evaluate the measurement imprecision with mean squared error, which,
\begin{equation}
\begin{aligned}
 \Delta \tilde{d}^2_\t{tot}= &  \t{E}\left[\left(\hat{d}-\tilde{d}\right)^2\right],\\
  =&\t{E}\left[\left(\hat{d}-\t{E}\left[\hat{d}\right]\right)^2\right]+\t{E}\left[\left(\t{E}\left[\hat{d}\right]-\tilde{d}\right)^2\right],\\
  =&\Delta \tilde{d}^2+\Delta \tilde{d}^2_\t{bias}.
\end{aligned}
\end{equation}
The first term is the estimator variance, describing the precision, and the second term is the estimator bias, describing the accuracy. We approximate the MLE with a linear estimator and a non-linear estimator. They will have different amount of bias over a dynamic range of interest. 

When one copy of the scene is available, e.g., when we only use a single temporal mode, the measurement outcomes are independent Poisson distributed random variables for all the transceivers under study here.
For this case, maximizing the log-likelihood function is equivalent to (using $Z$-mode coherent state transmitter and $\zeta$-SPADE receiver as an example), 
\begin{equation}
\begin{aligned}
    \hat{d}_{ML}(\vec{\gamma}_\mathrm{meas})=&\mathrm{Maxarg}\log\left[\prod\displaylimits_{j=0}^{1}P_{j}\left(\gamma_{j};\tilde{d}\right)\right]\\
    =&\mathrm{Maxarg} \sum\displaylimits^{1}_{j=0}\left[ -N_{j}+\gamma_{j} \log{N_{j}}-\log{\gamma_j!}\right],\\
    =&\mathrm{Maxarg} \sum\displaylimits^ {1}_{j=0}\left[ -N_{j}+\gamma_{j} \log{N_{j}}\right].
\end{aligned}
\label{MLE_meritfunc}
\end{equation}

The $N_j$'s are functions of $\tilde{d}$ and $\vec{\gamma}_\mathrm{meas}=(\gamma_0,\gamma_1)$
are not functions of $\tilde{d}$. Thus, we discard the third term in the optimization. When $v$ copies of the scene are available, e.g., when the number of temporal-spatial modes employed is $v\times 2$, the MLE becomes 
\begin{equation}
 \hat{d}_{ML}({}^{1}\vec{\gamma}_\mathrm{meas},{}^{2}\vec{\gamma}_\mathrm{meas},\cdots,{}^{v}\vec{\gamma}_\mathrm{meas})=\mathrm{Maxarg }\sum\displaylimits^{1}_{j=0} \left[-N_{j}+ \log{N_{j}}\sum_{l=1}^{v}\frac{{}^{l}\gamma_{j}}{v} \right],
 \label{eq_multicopy_MLE}
\end{equation}
where ${}^{l}\vec{\gamma}_\mathrm{meas}$ is the measurement outcome of the $l$-th temporal mode. In order to have a simple and fast estimator, we can approximate $N_j$
\begin{equation}
\begin{aligned}
    N_j=&\left| \iint_{-\infty}^{+\infty} \left[\sqrt{N \eta_Z} \underbrace{\zeta_0(x+d, y) A_\t{d}(x,y) }_{f(d)} \right]\zeta_j^*(x,y) \mathrm{d} x \mathrm{d}y\right|^2,\\
    =&N \eta_Z \left|\iint_{-\infty}^{+\infty}\left[\sum_{n=0}^{g-1} \frac{f^{(n)}(0)}{n !} d^n+ \mathcal{O}\left(d^g\right)\right] \zeta_j^*(x ,y) \mathrm{d} x \mathrm{d}y\right|^2.
\end{aligned}
\label{N_TSexpansion}
\end{equation}
The bracket in the first line of Eq.\eqref{N_TSexpansion} shows the received field when a coherent state is sent through the transmitter in $Z(x, y)$ mode with a total mean photon number $N$. 
Performing Taylor series expansion up to $d^{g-1}$ order on $f(d)$, we get an approximated $N_j$. Approximated $N_j$ leads to simplified estimator devised from MLE. 

When we take $g=2$, we have, $N_0=N\eta_Z$ and $N_1= N \eta_Z \beta \tilde{d}^2$. Applying Eq. (24) leads to a linear estimator of the form,
\begin{equation}
\hat{d}_L (\vec{\gamma}_\t{meas})=\sqrt{\frac{\gamma_1}{N\eta_Z \beta}},
\end{equation}
in which $\gamma_1$ is the measured photon number from mode $\zeta_1(x,y)$. When we take $g$ values higher than $2$, the NL estimator will have a more involved form. The approximated MLE merit function can be easily modified for other transceivers with the same procedure presented above. First, we linearize the MLE by taking $g=2$, then we increase $g$ for every transceiver design to get an unbiased approximated MLE over $\tilde{d}\in(0,0.01)$. All the simulation is done for a system with $D_\text{F}=90$ and presented in Fig. 4 of the main text.

 For a setup with $D_\text{F}=90$, over the total integration time, the transmitter consumed a total mean photon number, $N=10^{11}$, for each trial. $1000$ trials were ran for discrete values of $\tilde{d}\in(0,0.1)$. The total error of the linear estimator are bias dominated. The total error of the nonlinear estimator in $\tilde{d}\le0.01$ are shot noise limited with stable precision gap between the classical optimal transceiver design and the conventional design. The NL estimators for the four transceivers take $g$ values of $\{10,10,6,6\}$ respectively. With the NL estimator, we simulated the total estimator error for $\tilde{d}=\{0.001,0.01,0.1\}$ with different total mean photon numbers, $N=\{10^7,10^8,10^9,10^{10},10^{11}\}$. As shown in Fig. 4(b), when the total error is shot noise limited, it follows classical scaling, $\Delta d_\t{tot}\propto 1/\sqrt{N}$. The black dashed lines in Fig. 4(b) are the corresponding precision lower bound set by QFI and/or CFI for the corresponding transceivers in the small transverse displacement limit. 
 
%